\documentclass[11pt,twoside]{article}
\usepackage{./asp2014}
\aspSuppressVolSlug
\resetcounters

\def\etal{{\it et\thinspace al.}\ }
\def\eion{{(e~+~ion)}\ }
\def\efe18{{(e~+~Fe~XVIII)}\ }
\def\efe1817{{(e~+~Fe~XVIII) $\rightarrow$ Fe~XVII}\ }

\begin{document}

\title{Recalculation of Astrophysical Opacities: Overview, Methodology
and Atomic Calculations}
\author{Anil K. Pradhan \& Sultana N. Nahar}
\affil{The Ohio State University, Columbus, Ohio, USA 43210}
\email{pradhan.1@osu.edu}

\paperauthor{Anil Pradhan}{pradhan.1@osu.edu}{ORCID_Or_Blank}{Ohio State
University}{Astronomy}{Columbus}{Ohio}{43210}{USA}
\paperauthor{Sultana Nahar}{nahar.1@osu.edu}{ORCID_Or_Blank}{Ohio State
University}{Astronomy}{Columbus}{Ohio}{43210}{USA}

\begin{abstract}

 A review of a renewed effort to recalculate astrophysical opacities
using the R-Matrix method is presented. The computational methods and
new extensions are described. Resulting enhancements found in
test calculations under stellar interior conditions compared to the
Opacity Project
 could potentially lead to the resolution of the solar abundances
problem, as well as discrepancies between recent experimental
measurements at the
Sandia Z-pinch inertial confinement fusion
device and theoretical opacity models. Outstanding issues also discussed
are: (i)
accuracy, convergence, and completeness of atomic calculations,
(ii) improvements in the Equation-of-State of high-temperature-density
 plasmas, and 
(iii) redistribution of resonant 
oscillator strength in the bound-free continuum, and
(iv) plasma broadening of auotionizing resonances.

\end{abstract}

\section{Introduction}

 The Opacity Project was launched in 1983 with the goal of calculating
astrophysical opacities using state-of-the-art atomic physics
based on the coupled channel (CC) approximation employing the
powerful R-Matrix (RM) method (Seaton \etal 1994,
{\it The Opacity Project Team} 1995,
1996). Over the next decade, a suite of
extended RM opacity codes were developed
to compute large-scale bound-bound transition strengths and
bound-free photoionization cross sections with unprecedented accuracy
One of the primary features of OP was the precise delineation of
intrinsic autoionizing resonance profiles whose shapes, extent and 
magnitudes are
determined by myriad channel couplings in the (electron+ion) system.
However, CC-RM calculations are of immense complexity and require
substantial computational effort and resources. For the often 
dominant inner-shell
transitions they could not be completed owing to computational constraints on
the then available
high-performance supercomputing platforms. Simpler approximations akin
to distorted-wave (DW) type methods used in other opacity models 
that neglect channel couplings,
were therefore employed to compute most of the OP data. 

 In recent years a renewed effort has been under way
as originally envisaged using the CC-RM methodology (Nahar and Pradhan
2016a; hereafter NP16a), stimulated 
by two independent developments. The first was a 3D Non-LTE analysis of solar
elemental abundances that were up to 50\% lower for common volatile
elements such as C, N , O and Ne (Asplund \etal 2009). 
It was suggested that an enhancement
of up to 30\% in opacities could resolve the discrepancy, particularly in 
helioseismological models (Bahcall \etal 2005; Basu and Antia 2008;
Christensen-Dalsgaard \etal 2009; J. Bahcall, private communication). 
The second was an experimental
measurement of iron opacity at the Sandia Z-pinch inertial confinement
fusion device,
under stellar interior conditions prevalent at the
base of the solar convection zone (BCZ), that were 30-400\% higher in
monochromatic opacity compared to OP (Bailey \etal 2015). 
The Z-pinch results for the
Rosseland Mean Opacity (RMO) were also substantially higher using the measured
data, and found nearly half the enhanecment 
needed to resolve the solar abundance problem.

 The pilot CC-RM calculations in NP16 for
an important iron ion Fe~XVII resulted in 35\% enhancement relative to
the OP RMO at the Z conditions. While the enhancement
is consistent with
subsequently reported results from other opacity models (Blancard \etal
2016, Nahar and Pradhan 2016b), 
there are also important
differences in (i) atomic physics, (ii)
equation-of-state, and (iii) plasma broadening of autoionizing resonances.
The Fe~XVII calculations were carried
through to convergence by including n = 3 and n = 4 levels of the target
ion FeXVIII. They showed large enhancements in photoionization cross
sections, as successive thresholds are included, due to coupled
resonance structures and the background. The extensive role of 
photoionization-of-core (PEC) or Seaton resonances (Yu and Seaton 1987)
associated with strong dipole
transitions in the core ion Fe~XVIII is especially prominent (NP16a). Several
other sets of the pilot calculations have been carried out: 
relativistic Breit-Pauli R-Matrix (BPRM) 
calculations incluging 60 fine structure 
levels up to the n = 3 thresholds, non-relativistic calculations
including 99 LS terms up to the n = 4 threshold, as well as BPRM
calculations with 218 fine structure levels (in progress). 
 One of the aims is to
benchmark existing DW cross sections and 
monochromatic opacities in the non-resonant background and
the high energy region above all coupled excitaion thresholds. 

 Among the compromises necessary at the time of the intial OP
work related to very small wavefunction expansions for the \eion
system in R-Matrix calculations, usually limited to the ground
configuration of the core ion. One of the main points of the 
NP16a work was that {\it in general for any atomic system a converged
expansion in terms of the target configurations and levels is needed to
include the full enhancement of photionization cross sections for each
level in the bound free continuum}. 

We are also investigating
occupation probabilities from the Mihalas-Hummer-Dappen equation-of-state 
employed in the OP work which are orders of magnitude lower for excited
levels than other models. 

A new theoretical method and computational
algorithm for electron impact broadening of autoionizing resonances in
plasmas, as function of temperature and density, is described [9].
Finally, issues related to completeness and accuracy
are addressed [6,10]. 

 Following sections describe some of the salient features of the work
outlined above.

\section{Close coupling and distorted wave approximations}

   The Close-Coupling approximation is
implemented using the R-Matrix method (CC-RM). It
involves the expansion of the total
wavefunction for the (electron + ion) system in terms of the
eigenfunctions of the ``target" or the ``core" ion states and a free
electron wavefunction.

\begin{equation}
\Psi(E) = A \sum_{i} \chi_{i}\theta_{i} + \sum_{j} c_{j} \Phi_{j},
\label{eq:psi}
\end{equation}
where $\chi_{i}$ is the target ion wave function in a specific state
$S_{i}$
$L_{i}$ and $\theta_{i}$ is the wave function for the free electron in a
channel labeled as $S_{i}L_{i}k_{i}^{2}\ell_{i}(SL\pi)$; $k_{i}^{2}$
being its incident kinetic energy. In contrast to the CC-RM method, the
DW approximation, used in existing opacity models,
neglects the summation over hundreds to thousands of
channels on the RHS. In essence that implies that the DW method neglects
the quantum superposition and interference that gives rise to
autoionizing resonances in an {\it ab initio} manner.
 To obtain bound-bound, bound-free, and scattering matrix elements,
 we obtain (e+ion) wavefunctions $\Psi_B(SL\pi;E)$ and
$\Psi_F(SL\pi;E')$. For bound states  B and B'
the line strength in a.u. is given by
  $\Gamma(B;B') = |<\Psi_B(E_B)||{\bf D}||\Psi_{B'}(E_{B'})>|^2$,
where {\bf D} is the dipole operator. If the final state is a continuum
state represented by $\Psi_F(E')$ and the initial state by $\Psi_B(E)$
then the photoionization cross section is given by
 $\sigma_{\omega}(B;E') = \frac{4}{3}\frac{\alpha\omega}{g_i}
|<\Psi_B(E_B)||{\bf D}||\Psi_F(E')>|^2$
where $\omega$ is the photon frequency and $E'$ is the energy of the
outgoing electron.
The relativistic BPRM method incorporates the Breit-Pauli
 Hamiltomian for the (N+1)-electron system. We employ a
pair-coupling representation $S_iL_l(J_i)l_i(K_i)s_i (J\pi)$. As the
individual states $S_iL_i$ split into the fine structure levels
$J_i$, the number of channels becomes several times larger than the
corresponding LS coupling case. Hitherto, the version of BPRM codes
included only the 1-body terms of the Breit interaction, namely
the spin-orbit (so) coupling, mass correction (m), and Darwin (d)
terms (Eissner 1991).
Chen and Eissner (to be published) have developed an
improved Breit-Pauli version including the 2-body terms in the Breit
interaction (Fig.~1),
that should be more accurate for fine structure and
resonance effects in the Fe-group elements.
 
\section{R-Matrix Opacity Codes}
 The R-Matrix opacity codes (Seaton 1987, Berrington \etal 1987) are
significantly different from the original R-Matrix codes (Burke 2011),
and have been considerably extended by the OSU group for complete RM
opacity calculations (Fig.~1, Fig.~2).
Fig.~1 is a flow-chart of the RM codes set up and used
by the PI
and collaborators at the Ohio Supercomputer Center (OSC) since 1990.
An accurate
configuration interaction representation or the core ion
states is obtained by two atomic structure codes, {\bf SUPERSTRUCTURE}
(Eissner
\etal 1974) and {\bf CIV3} (Hibbert 1975).
The first two R-matrix codes, {\bf STG1,STG2}, are then employed to
generate the multipole integrals, algebraic coefficients and set up
the (N+1)-electron Hamiltonian corresponding to the coupled
integro-differential equations. The Hamiltonian is diagonalized in
{\bf STGH}; in the BP calculations the diagonalization is preceded by
LSJ recoupling in {\bf RECUPD}.
The R-matrix basis set of functions and the dipole matrix elements
so produced are then input into {\bf STGB} for bound state
wavefunctions,
{\bf STGF} for continuum wavefunctions, {\bf STGBB} for radiative
transition probabilities, and {\bf STGBF} for photoionization cross
sections.  In addition, {\bf STGF(J)} is used to obtain
collision strengths for electron impact excitation in LS or
intermediate coupling and fine structure transitions.

\articlefigure[width=.7\textwidth,height=.4\textheight]{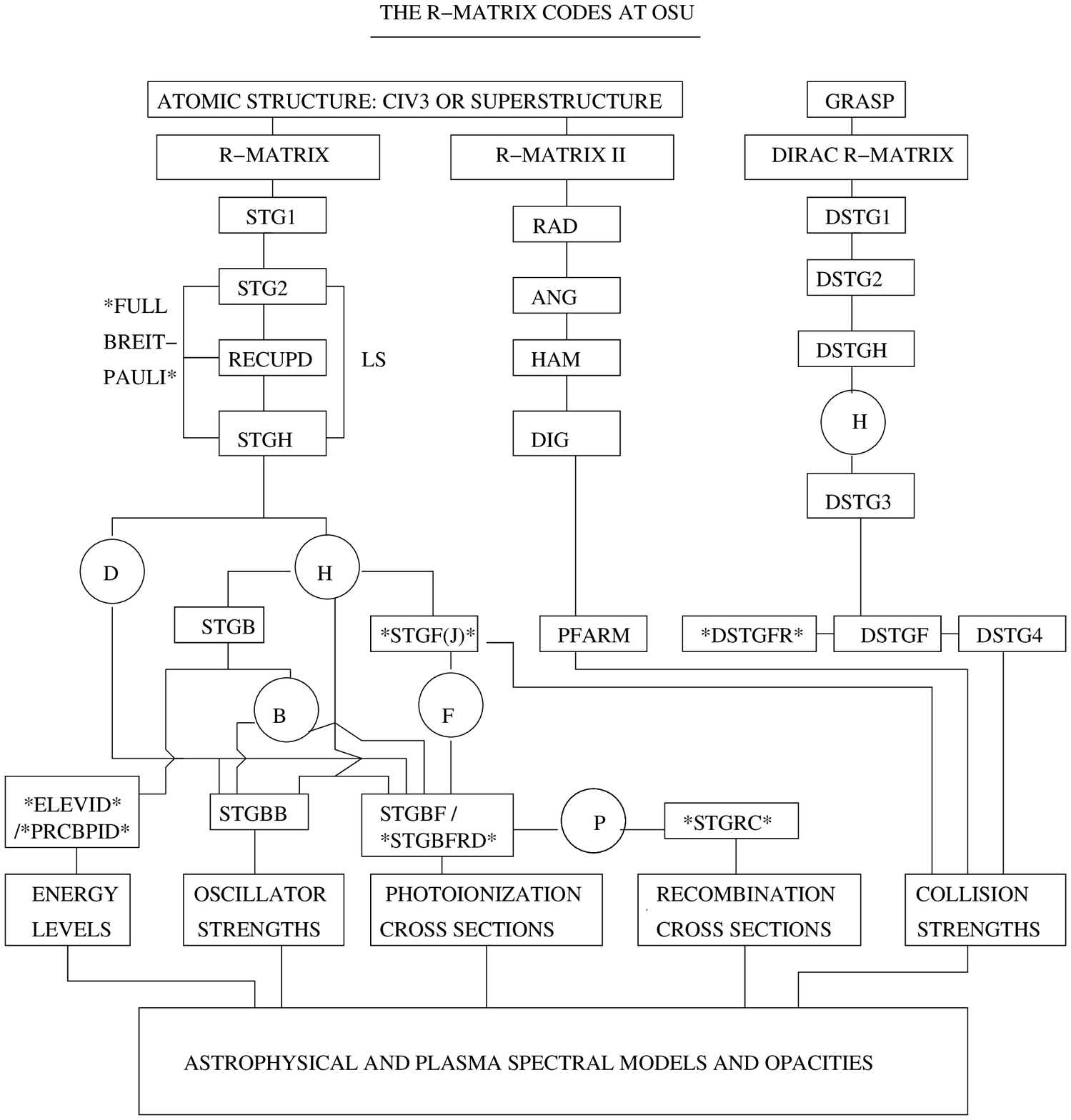}{ex_fig1}{The
R-Matrix codes for calculation of opacities. The left branch of
Breit-Pauli codes is employed in the results presented in this work.}

  The flowchart of the new package of codes for opacities calculations
is shown schematically in Fig.~1;
each of the boxes
represents a series of subsidiary codes.
The new version of the R-Matrix
opacities codes is the extension of following codes under development:
(A) INTFACE --- to interface atomic
data for bound-free photoionization and bound-bound
transition probabilities including relativistic fine structure, and (B)
AUTOBRO --- an algorithm for electron impact broadening of
autoionizing resonances with a temperature-density dependent kernel,
HIPOP --- high-precision version of the opacities
code with high-resolution
($10^5$ frequencies) and monochromatic opacities and Rosseland and
Planck
means for arbitrary mixtures. The database OPSERVER is described by
Mendoza \etal (2007); its updated version has been established at CDS
by F. Delahaye.

\articlefigure[width=.5\textwidth]{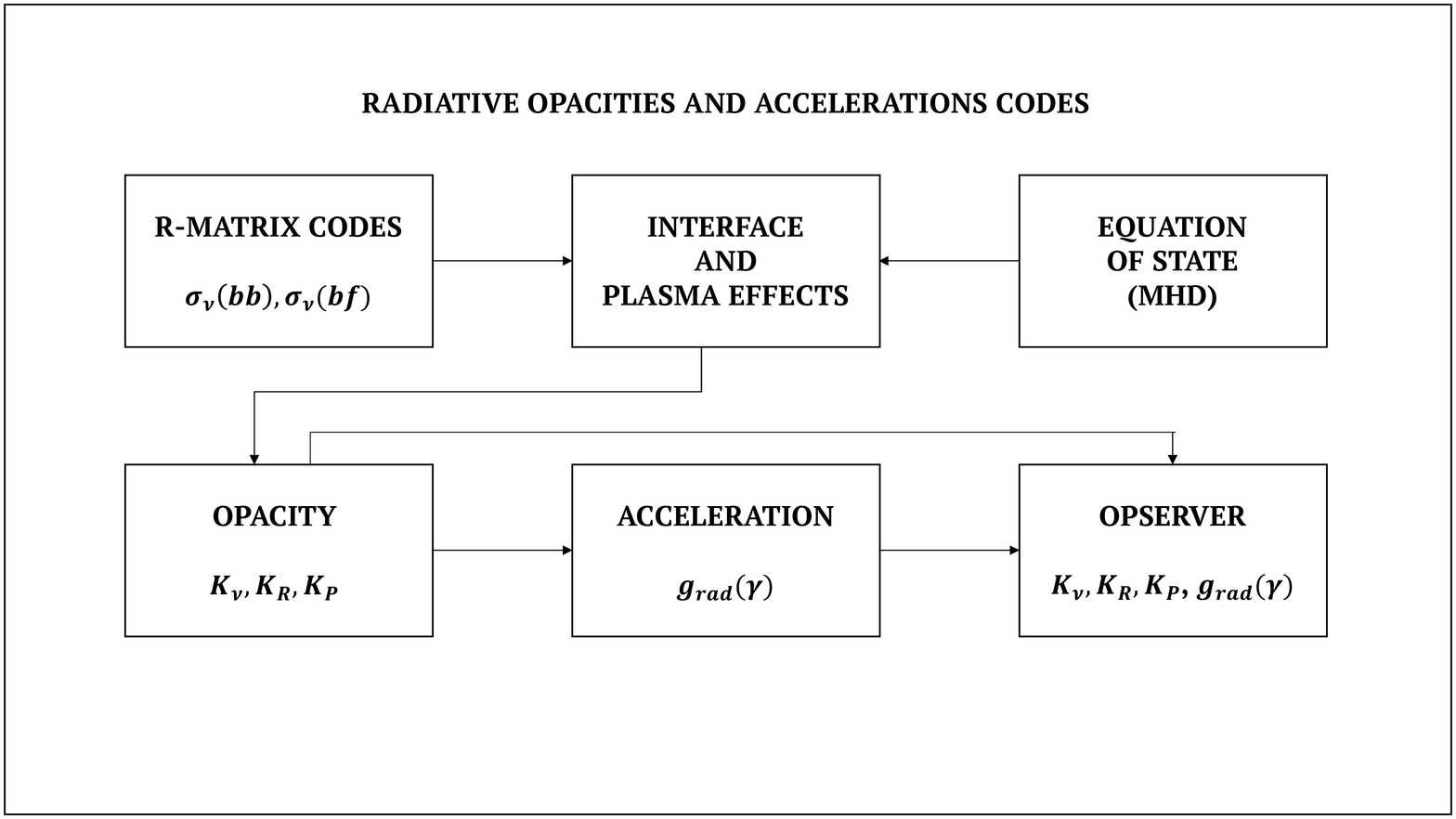}{ex_fig2}{The opacity
codes including interfacing with CC-RM data and the
equation-of-state calculations.}

\section{Inner- and outer-shell excitations}

 The CC-RM approach entails an increasing
succession of open channels as each threshold of the core ion is
approached from below (Eq.~1). As the number of free channels (open +
closed) multiplies, the sizes of the \eion algebra and R-Matrix 
diagonalization matrices and arrays may become very large. In addition, there
are issues related to spectroscopic identification of a highly excited
and mixed-configuration bound
\eion levels from STGB calculations,
 and subsequent
bound-bound STGBB and bound-free STGBF calculations (Fig.1). While
these have been solved, they require commensurate computational
resources and expertise.

 Inner-shell excitations using the DW method have been described in
earlier works (e.g. Badnell and Seaton 2003).
Despite that fact that most of current OP data was computed using
variants of DW data, the initial OP calculations were done using the RM
method, albeit with relatively small eigenfunction expansion and only
few outer-shell excitations. Most of OP codes were
originally developed to employ the RM methodology,
but computational constraints precluded OP work
from incorporating resonant inner-shell atomic processes that
manifest themselves strongly in the {\em bound-free} cross sections.
But inner-shell excitations are
dominant contributors to opacity because most electrons
in complex ions are in closed shells, whose
excitation energies lie above the (first) ionization threshold. Much of
the opacity therefore lies in the strongest such transitions that are
associated with dipole transitions in the core of the ion. The
corresponding
autoionizing resonances are referred to as {\it Photoexciation-of-Core}
(PEC) or Seaton resonances (Yu and Seaton 1987, Nahar \etal 2011, NP16;
and discussed extensively in {\it Atomic Astrophysics and
Spectroscopy}, Pradhan and Nahar 2011).
{\it The PEC resonances are the largest single contributing feature to
the
bound-free opacity.} Therefore, their simple
treatment as bound-bound transitions
in current DW opacity models needs verification.

\section{New R-Matrix opacity calculations: Test case of Fe~XVII}
 
  The monochromatic opacity comprises of bound-bound (bb),
bound-free (bf), free-free (ff) and photon scattering (sc)
contributions:

\begin{equation}
 \kappa_{ijk}(\nu) = \sum_k A_k \sum_j f_j \sum_{i,i'}
[\kappa_{bb(}(i,i';\nu) +
\kappa_{bf}(i,\epsilon i';\nu) + \kappa_{ff} (\epsilon i, \epsilon' i';
\nu) + \kappa_{sc} (\nu)] ,
\label{eq:k}
\end{equation}

 where $A_k$ is the abundance of element $k$, $f_j$
the ionization fraction $j$, $i,i'$ are the initial bound and
final bound or continuum states of the atomic species, and $\epsilon$
represents the electron energy in the continuum.
 The Rosseland Mean Opacity (RMO) $\kappa_R$ is defined in terms of
$\kappa_{ijk}(\nu)$ as

\begin{equation}
 \frac {1}{\kappa_R} = \frac{\int_0^\infty g(u) \frac{1}{\kappa_\nu}
du}{\int_0^\infty g(u) du}, \ \ \ g(u) = u^4 e^{-u} (1 - e^{-u})^{-2},
\label{eq:RMO}
\end{equation}

where $g(u)$ is the derivative of the
Planck weighting function (corrected for stimulated
emission), $\kappa^{bb}_\nu (i \rightarrow j) = \left( \frac{\pi
e^2}{m_e
c} \right) N_i f_{ij} \phi_\nu$, and $\kappa^{bf}_\nu = N_i \sigma_\nu$.
The $\kappa_\nu$ is primarily a function of the {\em bb} oscillator
strengths
$f$, {\em bf}  photoionization cross sections $\sigma_\nu$, level
populations
$N_i$, and the line profile factor $\phi_\nu$.

 The CC-RM computational framework for large-scale computations
comprises mainly the first two components of the opacity
on the RHS of Eq.~\ref{eq:k} (i)
the bound-bound (bb) transition probabilities, and (ii) photoionization
of bound-free (bf) cross sections. 

 We focus on test calculations for an important contributor to opacity
at the Z-pinch plasma conditions: $T = 2.11 \times 10^6$K and $N_e = 3.1
\times 10^{22}$ cc; the experimental energy range and the range of the
Planck function derivative dB/dT is shown in Fig.~3 (Right). 
The Fe~XVII calculations involved
 coupled channel wavefunction expansion in Eq.~\ref{eq:psi}
including 30 LS terms with 60 fine structure levels up to the $n \leq 3$
complex of the core ion Fe~XVIII (Nahar \etal 2011),
and 99 LS terms or 218 levels up to $n \leq 4$ (NP16a);
the latter calculation was an
order of magnitude larger computationally.
The associated $n = 2 \rightarrow 3$ core transitions in Fe~XVIII in the
former case, and $n = 2 \rightarrow 3,4$ in the latter case,
give rise to numerous PEC resonances in
the bound-free photoionization cross sections of Fe~XVII. Fig.~3 (Left) from
NP16a compares enhancement in the new cross sections compared to 2-state
coupled channel results from OP that reproduce only the background.
There is no "misrepresentation" of OP as asserted by Iglesias and Hansen
(2017; hereafter IH17), since that is the only other CC
calculation available; rest of the OP work is DW.

\articlefiguretwo{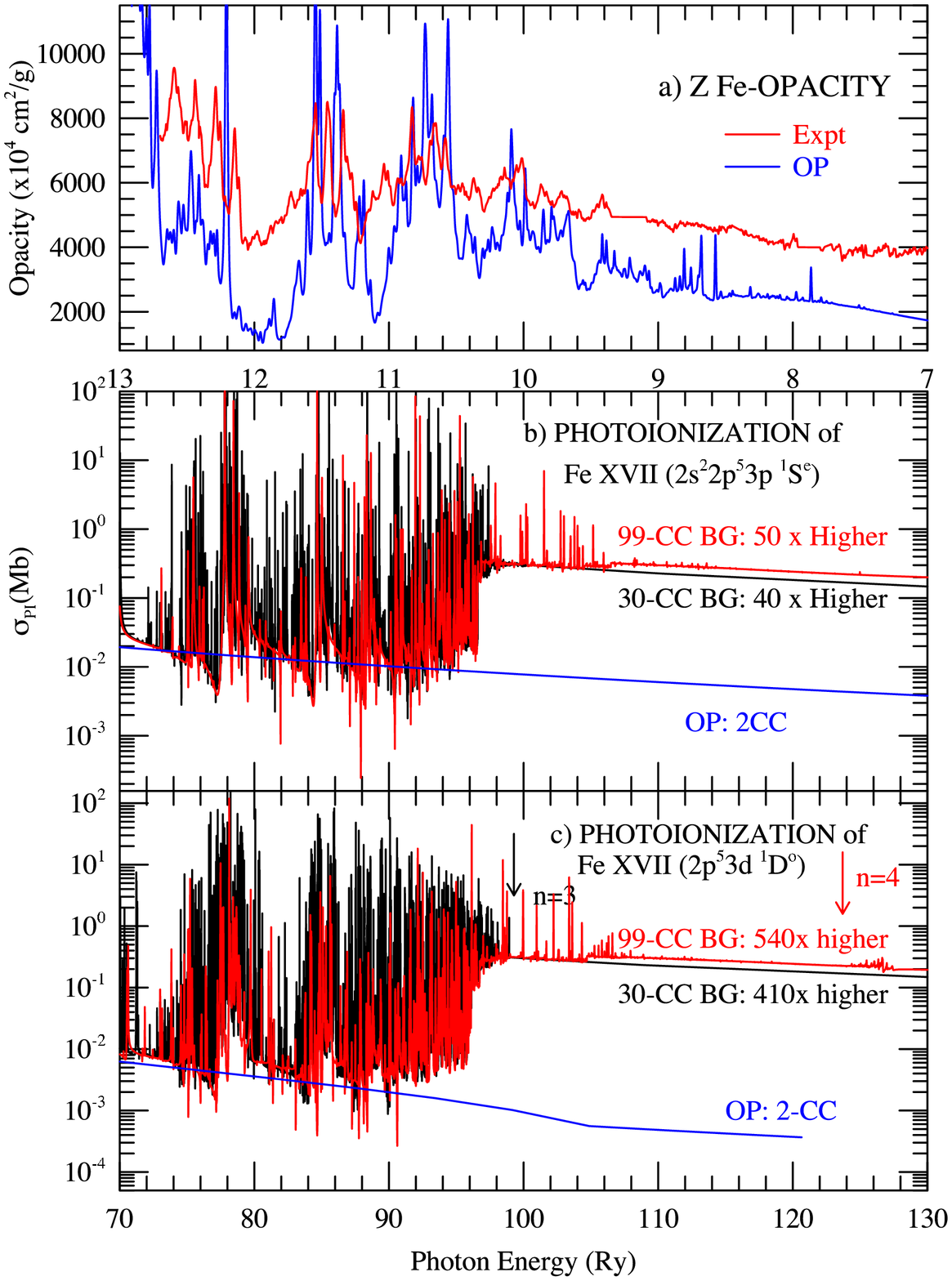}{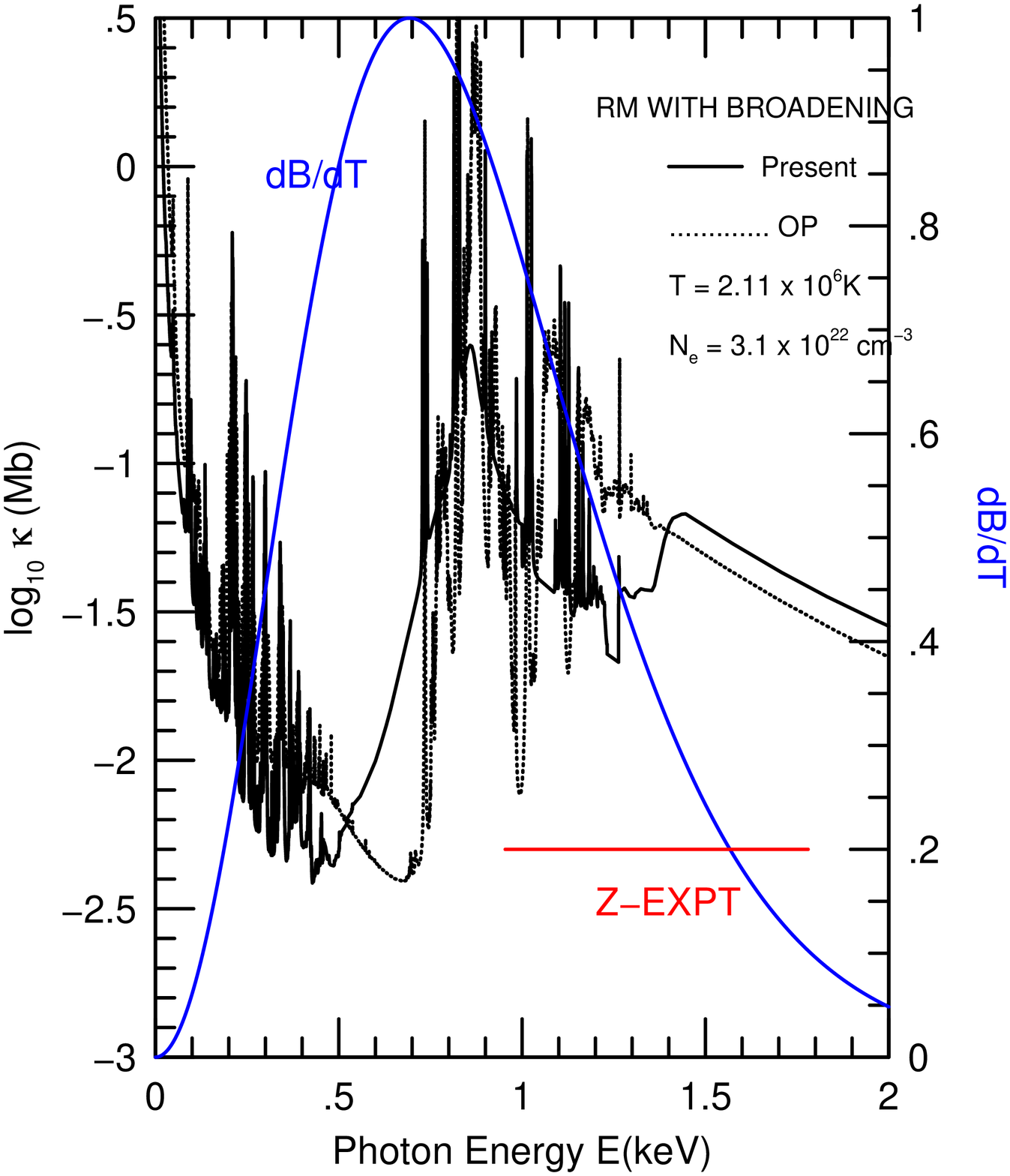}{ex_fig3}{\emph{Left}: Autoionizing
resonance structures in R-Matrix
photoionization cross sections of two Fe~XVII excited levels (bottom and
middle panels); "higher-than-predicted" Sandia Z-pinch measurements of
total Fe opacity (top panel, Bailey \etal 2015). Calculations are needed
for Fe~XVI-XXI to compare with experiments at Z plasma conditions.
\emph{Right:} Monochromatic R-Matrix opacity of Fe~XVII at
T = $2.11 \times 10^6$K and N$_e = 3.1 \times 10^{22}$
cm$^{-3}$, corresponding to total iron opacity measurements at the
Sandia Z-pinch fusion device in the energy range shown by
the red bar (Fig.~3, Left); the R-Matrix Fe~XVII RMO is
$\sim$60\% higher than the OP (Table~1).}

\section{Convergence and completeness}

The two criteria for {\it accuracy and completeness} of
RM opacity calculations are: (I) {\bf
convergence} of
wavefunction expansion (Eq.~3) with respect to
bound-free cross sections and bound-bound transition
probabilities, and
(II) {\bf completeness} of monochromatic and RMOs
with respect to a very large number of possible
multiply excited configurations that contribute to the
background non-resonant cross sections
in high-temperature plasmas. The NP16a work demonstrated convergence
with respect to
the Fe~XVIII target ion included in the CC-RM calculations (Fig.~3,
Left), but the highly excited configurations were not included that
affect the high-energy behavior.

Although the number of these excited configurations is large, their 
contribution to opacity is still
small relative to the main CC-RM cross sections. It is also specific to
the energy range where the quasi-degenerate configurations lie.
In addition, they represent the background contribution without
resoances; thus amenable to simpler approximations such as the DW
without loss of accuracy.
We refer to these configurations as 
"top-up" contribution to the CC-RM calculations. 

 Whereas enumerating excited configurations is straightforward in atomic
structure-DW calculation (IH17 and  Zhao \etal in this
volume), it is more complex and indirect in CC-RM calculations owing to
resonant phenomena (IH17 incorrectly state that some of the
configurations were omitted in NP16a; in fact the $nl = 4d, 4f$ are included). 
For example, the 60CC n$\leq3$ BPRM calculations
yield 454 bound levels of Fe~XVII, but we have further included
$>50,000$ topup levels to compute the opacity spectrum in Fig.~3
(Right). However, the photoionization cross sections of the 454 strictly bound
levels (-ve energy eigenvalues) includes embedded autoionizing
resonances that are treated as distinct levels in the DW calculations.
Therefore, in total there are commensurate number of levels
to ensure completeness.
 Zhao \etal discuss the complementary topup to the CC-RM calculations
and find a $\sim$20\% increment.
Satisfying both
these criteria results in a further enhancement in the Fe~XVII RMO over those
obtained in NP16a of $\sim60$\% over OP. 
Table~1 shows the latest CC-RM results compared
to other models. We expect additional enhancement upon
completion of the n=4 fine
structure calculations, that is likely the cause of the deficit in monochromatic
opacity just below 1.5 keV in Fig.~3 (Left); 
updated results will be presented in Pradhan \etal (2017). 

\begin{table}
\begin{center}
\scriptsize
\caption{Comparison of recent R-Matrix results (Pradhan \etal 2017)
with other
opacity models for Fe~XVII employing variants of the DW method (Blancard
\etal 2016), are over 60\% higher;
further enhancement in R-Matrix opacities
 is expected after proposed improvements in more
extended atomic calculations and the MHD-EOS.}
\vspace{0.1in}
\begin{tabular}{ll}
\hline
\noalign{\smallskip}
 Opacity Enhancement Factors for Fe~XVII relative to OP\\
\hline
\noalign{\smallskip}
 R-Matrix (Pradhan \etal 2017) & 1.64\\
 OPAS (Blancard \etal 2012) & 1.55\\
 SCO-RCG (Blenski \etal 2000)& 1.37\\
 ATOMIC  (Magee \etal 2013) & 1.32\\
 SCRAM (Hansen \etal 2007) & 1.27\\
 TOPAZ  (Iglesias 2015) & 1.21\\
\hline
\end{tabular}
\end{center}
\end{table}

\section{Differential oscillator strength distribution and sum-rule}

 The oscilltor strength sum-rule is invoked to ensure completeness. The
summation over all bound-bound and bound-free transitions must satisfy
the condition built into the definition of the oscillator
strength as a fractional probability of excitation; Ergo: $\sum_j f_{ij}
= N$, where N is the number of active electrons. 
However, while the $f$-sum rule ensures completeness, it 
does not relate to the accuracy of atomic
calculations per se. Rather, it is the precise energy distribution of 
differential oscillator strength that is the determinant of accuracy.
To wit: a hydrogenic approximation for complex atoms would satisfy the
$f$-sum rule, but it would clearly be inaccurate. As demonstrated above,
the CC-RM method is concerned primarily with differential oscillator
strength in the bound-free continuum that, in turn, depends on
delineation of auotionizing resonances.

 Similarly, the RMO depends on the energy distribution monochromatic
opacity via the Planck function at a given temperature.
Fig.3 (Right) shows dB/dT at the Z-pinch temperarture of 2.11 $\times
10^6$K. Compared with the OP results, the distribution of the CC-RM Fe~XVII
monochromatic opacity is quite different and much more smoothed out,
without the sharp variations that stem maily from the treatment of
resonances as bound-bound lines (albeit inclusion of limited
autoionization broadening perturbatively in the DW approximation. The
flatter distribution is also observed experimentally, in contrast to
theoretical opacity models that exhibit "opacity windows" (Fig.3, Left).

\section{Plasma broadening of autoionizing resonances}

 An unsolved atomic physics problem is the broadening of autoionizing
resonances by plasma effects, line broadening theory which is well
developed and extensively
considered in existing opacity calculations. But since autoionizing
resonance shapes are not considered, neither is their broadening as
function of density and temperature. However, the distribution of
differential oscillator strength and the structure of the bound-free
opacity depends critically on how resonances broaden and
dissolve into the continuum from the CC-RM calculations.

 At high densities the dominant form of plasma broadening is due to
electron impact. We have developed an algorithm implemented in the
INTFACE code (Fig.~2), prior to opacity calculations (Pradhan 2017). 
Fig.~4 shows
electron impact broadening of autoinizing resonances in a typical CC-RM
photoionization cross section at two
temperatures, 10$^6$ and 10$^7$K, and a range of densities:
$N_e = 10^{20}$ cc where the onset of broadening is discernible, to high
densities $N_e = 10^{23}$ cc where the resonance structures are mostly
dissolved into the continuum. The broadening profiles are normalized
Lorentzian, with no effect on the non-resonant continuum, viz. Fig.~3
(Right). 

\articlefigure[width=.7\textwidth,height=.4\textheight]{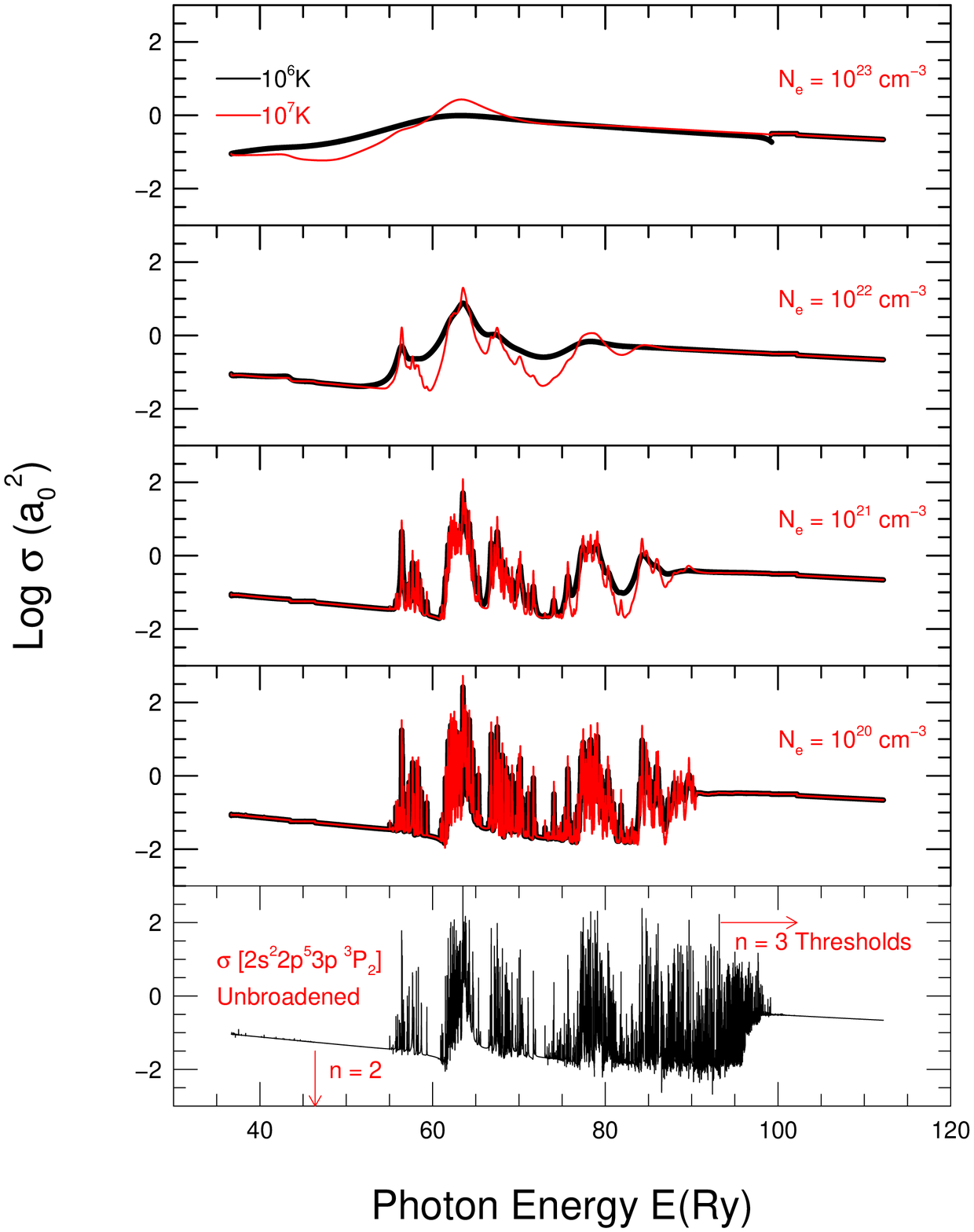}{ex_fig4}{Plasma broadening
of autoionizing resonances with temperature and
densities, bottom panel --- unbroadened cross section, upper panels ---
dissolution of resonance structures into the bound-free continuum with
increasing density at T = 10$^6$K (black) and 10$^7$K (red).}


\section{Equation-of-State}

The Mihalas-Hummer-D\"{a}ppen equation-of-state (MHD-EOS) was formulated
for OP (Mihalas \etal 1988; {\it The Opacity Project Team} 1995). 
The MHD "chemical picture" treats isolated atoms, for which the CC-RM
data are computed, perturbed by plasma environment via occupation
probabilities $w_i$ in the atomic internal partition function

\begin{equation}
 U(i,j,k) = \sum_i w_i g_i exp(-E_i/kT),
\label{eq:eos}
\end{equation}.


 and the modified Boltzmann-Saha equation; 
$E_i$ is the excitation energy of level $i$, $g_i$ its statistical
weight, and $T$ the temperature. The $w_i$ are determined upon
free-energy minimization of the plasma at a given temperature-density. 
An atomic level $i$ is considerd dissolved by
the plasma microfields
when its highest Stark sub-level overlaps with the lowest sub-level of
the $i+1$ level. The MHD-EOS originally
stipulated a stringent validity criterion for mass densities
$\rho \leq 0.01$ g/cc valid for stellar envelopes (Seaton \etal 1994).
But that is insufficient even to reach
the solar BCZ depths $rho \geq 0.1$ g/cc. Nevertheless, a modified
version named Q-MHD it has been
employed in the OP work, with $W_i$ cut-offs at very high densities
An unresolved problem
has also been that the chemical picture level populations are orders of
magnitude lower (Badnell and Seaton 2003)
than the physical picture EOS utilized in the Livermore
opacity calculations (Rogers and Iglesias 1992). 
An improved treatment 
of these issues in the
high-density regime has been developed by Trampedach
\etal (2006), and would be incorporated in the new RM opacity
calculations.

 Another issue pertains to topup levels referred to above, in addition
to the CC-RM levels. Including a large number of levels in the MHD-EOS
drastically affects the overall population distribution since it is
normalized via the internal partition function (Eq.~4). To wit: 
the ground state of Fe~XVII ends up with no more than a few percent of
the population at Z-pinch plasma and BCZ conditions. It is expected that
the improved method by Trampedach \etal would also alleviate this
problem.

\section{Conclusion}

  The R-Matrix opacity calculations described in this review are difficult and
time consuming, but ensure high accuracy using state-of-the-art atomic
physics, as originally envisioned in the OP. 
Furthermore, the CC-RM calculations are needed to verify 
existing opacity models that employ
some variant of the simpler DW method, but with large differences of
$\sim$30\% or more among them (viz.
Table~1). Ongoing and planned laboratory
experiments at the Sandia Z-pinch and the Livermore National Ignition
Facility (T. Perry in this volume) should also serve to validate theoretical
models to resolve outstanding discrepancies and fundamental issues in
atomic physics, astrophysics and plasma physics.

\acknowledgements{We would like to thank Werner Eissner, Regner
Trampedach, Lianshui Zhao and Chris Orban for contributions.
This work was partially supported by the U.S. Department of
Energy and the U.S. National Science Foudnation. The computational work
was carried out primarily at the Ohio Supercomputer Center in Columbus,
Ohio.}


\def\aa{{\it Astron. Astrophys.}\ }
\def\aasup{{\it Astron. Astrophys. Suppl. Ser.}\ }
\def\adndt{{\it Atom. data and Nucl. Data Tables.}\ }
\def\aj{{\it Astron. J.}\ }
\def\apj{{\it Astrophys. J.}\ }
\def\apjs{{\it Astrophys. J. Supp. Ser.}\ }
\def\apjl{{\it Astrophys. J. Lett.}\ }
\def\baas{{\it Bull. Amer. Astron. Soc.}\ }
\def\cpc{{\it Comput. Phys. Commun.}\ }
\def\jpb{{\it J. Phys. B}\ }
\def\jqsrt{{\it J. Quant. Spectrosc. Radiat. Transfer}\ }
\def\mn{{\it Mon. Not. R. astr. Soc.}\ }
\def\pasp{{\it Pub. Astron. Soc. Pacific}\ }
\def\pra{{\it Phys. Rev. A}\ }
\def\prl{{\it Phys. Rev. Lett.}\ }

\bibliography{ms}

\begin{thebibliography}{}
\bibitem[{{Asplund}(2009)}]{a09} M. Asplund, N. Grevesse, A. J. Sauval and P. Scott, {\it
Ann. Rev.  Astro. Astrophys.} 47, 481 (2009).
\bibitem[{{Badnell}(2003)}]{b03} N.R. Badnell and M.J. Seaton, \jpb, 36, 4367 (2003).
\bibitem[{{Bahcall}(2005)}]{b05} J.N. Bahcall, S. Basu, M.H. Pinsonneault, A.M. Serenelli,
\apj, 618, 1049 (2005).
\bibitem[{{Bailey}(2015)}]{b15}  J. E. Bailey, T. Nagayama, G. P. Loisel, G. A. Rochau,
C. Blancard, J.  Colgan, Ph. Cosse, G. Faussurier, C. J. Fontes, F.
Gilleron, I. Golovkin, S. B. Hansen, C. A. Iglesias, D. P. Kilcrease, J.
J.
McFarlane, R. C. Mancini, S. N. Nahar, C. Orban, J.-C. Pain, A. K.
Pradhan, M. Sherill \& B. G. Wilson, Nature, 517, 56-59 (2015).
\bibitem[{{Basu}(2008)}]{b08}  S. Basu and H.N. Antia, Physics Reports, 457, 217 (2008).
\bibitem[{{Blancard}(2016)}]{b16} C. Blancard, J. Colgan, Ph. Cosse, G. Faussurier, C.J. Fontes, F.
Gilleron, I. Golovkin, S.B. Hansen, C.A. Iglesias, D.P. Kilcrease, J.J.
MacFarlane, R.M. More, J.-C.Pain, M. Sherrill and B.G. Wilson,
\prl 117, 249501 (2016).
\bibitem[{{Burke}(2011)}]{b11}  P.G. Burke, {\it R-Matrix Theory of Atomic Collisions},
Springer Series on Atomic, Optical and Plasma Physics (2011).
\bibitem[{{Christensen-Dalsgaard}(2009)}]{c09}  J. Christensen-Dalsgaard, M. P. Di Mauro, G. Houdek, and F. Pijpers, \aa, 494, 205 (2009).
\bibitem[{{Eissner}(1974)}]{e74} W. Eissner, M.Jones, H. Nussbaumer \cpc 8, 270 (1974).
\bibitem[{{Iglesias}(2017)}]{i17}  C.A. Iglesias and S.B. Hansen, apj 835, 284 (2017).
\bibitem[{{Mendoza}(2007)}]{m07} C. Mendoza, M.J. Seaton, P. Buerger, 
A. Bellorin, M. Melendez, J.
Gonzalez, L.S. Rodriguez, E. Palacios, A.K. Pradhan, C.J. Zeippen, \mn,
378, 1031 (2007).
\bibitem[{{Mihalas}(1988)}]{m88}  D. Mihalas, W. D\"{a}ppen, and D.G. Hummer, \apj, 331, 815 (1988).
\bibitem[{{Nahar}(2011)}]{n11}  S.N. Nahar, A.K. Pradhan, W.Eissner and G.X. Chen,
Phys. Rev. A, 83, 053417 (2011).
\bibitem[{{Nahar}(2016)}]{n16a}  S.N. Nahar and A.K. Pradhan, \prl, 116, 235003 (2016).
\bibitem[{{Nahar}(2016)}]{n16b}  S.N. Nahar and A.K. Pradhan, \prl 117, 249502 (2016).
\bibitem[{{Pradhan}(2011)}]{b11}  Anil K. Pradhan and Sultana N. Nahar, {\it Atomic Astrophysics and
Spectroscopy}, Cambridge University Press (2011).
\bibitem[{{Pradhan}(2017)}]{p17}  A.K. Pradhan, S.N. Nahar, L. Zhao, W. Eissner, R.
Trampedach and C. Orban (2017, in preparation).
\bibitem[{{Rogers}(1992)}]{r92}  F.J. Rogers and C.A. Iglesias, \apjs, 79, 507 (1992).
\bibitem[{{Seaton}(1994)}]{s94}  M.J. Seaton,Y. Yu, D. Mihalas and A.K. Pradhan, \mn,266,805
(1994).
\bibitem[{{Opacity}(1995)}]{op95}  The Opacity Project Team, {\it The Opacity Project} Vol.1
(1995);
Ibid. Vol. 2 (1997), IOP Publishing, Bristol.
\bibitem[{{Trampedach}(2006)}]{t06}  R. Trampedach, W. D\"{a}ppen, V.A. Baturin, \apj, 646, 560 (2006).
\bibitem[{{Yu}(1987)}]{y87}  Y. Yu and M.J. Seaton, \jpb 20, 6409 (1987).

\end{thebibliography}
\end{document}